\documentstyle[aps]{revtex}   

\begin{document}
\draft
\title{ Interior Weyl-type Solutions of the Einstein-Maxwell Field Equations }
\author{Brendan S. Guilfoyle \footnote{Email: brendan.guilfoyle@ittralee.ie} }
\address{Mathematics Department, Institute of Technology Tralee, Tralee, Co. Kerry, Ireland.}
\date{\today}
\maketitle
\begin{abstract}
Static solutions of the electro-gravitational field equations exhibiting a functional relationship between the electric and gravitational potentials are studied.  General results for these metrics are presented which extend previous work of Majumdar.  In particular, it is shown that for any solution of the field equations exhibiting such a Weyl-type relationship, there exists a relationship between the matter density, the electric field density and the charge density.  It is also found that the Majumdar condition can hold for a bounded perfect fluid only if the matter pressure vanishes (that is, charged dust).  By restricting to spherically symmetric distributions of charged matter, a number of exact solutions are presented in closed form which generalise the Schwarzschild interior solution.  Some of these solutions exhibit functional relations between the electric and gravitational potentials different to the quadratic one of Weyl. All the non-dust solutions are well-behaved and, by matching them to the Reissner-Nordstr\"{o}m solution, all of the constants of integration are identified in terms of the total mass, total charge and radius of the source. This is done in detail for a number of specific examples.  These are also shown to satisfy the weak and strong energy conditions and many other regularity and energy conditions that may be required of any physically reasonable matter distribution.
\end{abstract}

\newtheorem{thm}{Theorem}[section]   
\newtheorem{cor}[thm]{Corollary}   
\newtheorem{lem}[thm]{Lemma}   
\newtheorem{prop}[thm]{Proposition}   
\newtheorem{defn}{Definition}[section]      
\newtheorem{rem}{Remark}[section]   
\newcommand{\nc}{\newcommand}
\nc{\sigf}{\Sigma_\phi}
\nc{\al}{\alpha}
\nc{\om}{\omega}
\nc{\bg}{\begin{eqnarray}}
\nc{\ed}{\end{eqnarray}}
\nc{\bp}{\bar{\Psi}}
\nc{\bm}{\bar{m}}
\nc{\ul}{\underline}
\nc{\bee}{\begin{equation}}
\nc{\ee}{\end{equation}}
\nc{\beer}{\begin{eqnarray}}
\nc{\eer}{\end{eqnarray}}
\nc{\M}{$(M,g_{ab})\:$}
\nc{\V}{$(V^3,g_{ij})\:$}
\nc{\lap}{\triangleup_2}
\nc{\dv}{\textstyle\frac}
\nc{\sdv}{\scriptstyle\frac}
\nc{\hf}{\dv{1}{2}}
\nc{\qtr}{\dv{1}{4}}
\nc{\shf}{\sdv{1}{2}}
\nc{\nn}{\nonumber}
\nc{\2}{\;\;}
\nc{\3}{\;\;\;}
\nc{\4}{\;\;\;\;}
\nc{\5}{\;\;\;\;\;}
\nc{\6}{\;\;\;\;\;\;}

\tableofcontents
\newpage

\section{Introduction}
The static field has played a central role in our understanding of Einstein's theory of gravitation.  Not only does this area lie closest to Newtonian gravitation and classical potential theory, but the existence of a timelike hypersurface orthogonal Killing vector brings about many simplifications in the field equations which allow one to integrate them exactly in numerous situations.  

In particular, the $g_{44}$ component of the metric tensor plays the part of a gravitational potential.  In the presence of an electric field a similar electric potential $\phi$ exists, and it was by considering fields in which there is a functional relationship between the two potentials that Weyl \cite{weyl} discovered a class of axially symmetric solutions.  We will refer to any static solution of the field equations with such a functional relationship as being of {\it Weyl-type}.  

Weyl considered only pure electric fields (that is, fields in which the electric field is the only source of the energy momentum tensor).  For these fields Weyl found that, in the presence of axial symmetry, the functional relationship, if it exists, must be of the form
\bee\label{e:weylrel}
g_{44}=A+B\phi+\phi^2.
\ee

Later work by Majumdar \cite{maj} showed that, even in the absence of axial symmetry, this quadratic restriction persists.  Majumdar also discovered that, if this quadratic relationship is a perfect square
\bee\label{e:majrel}
g_{44}=\left(C+\phi\right)^2,
\ee
then the spatial sections are conformally flat and one can interpret the space-time as describing the exterior field of a static ensemble, where the gravitational attraction and electric repulsion exactly balance.  These pure electric fields were independently discovered by Papapetrou \cite{pap1}.  We will refer to any static solutions of the field equations with a perfect square functional relationship (\ref{e:majrel}) as being of {\it Majumdar-type}.

The case of Weyl-type fields in which there are sources in addition to an electric field has received some attention.  Majumdar discovered solutions which represent the interior of a distribution of charged dust, held in balance by gravitational and electric forces.  Here again the functional relationship is a perfect square (\ref{e:majrel}) and the matter pressure vanishes, while the matter density is numerically equal to the charge density.  In addition, the total mass and total charge of the space-time are found to be numerically equal. 

Other than work done on charged dust, where the matter pressure vanishes (for example Das \cite{das}, G\"{u}rses \cite{gurs1} \cite{gurs2} Raychaudhuri \cite{ray}) little has been done to investigate Weyl-type fields in the presence of matter.  The question we address in this paper is whether the functional relation in the presence of matter sources (in particular, matter pressure) must be Weyl's quadratic one (\ref{e:weylrel}).  The answer to this is no, and we exhibit physically reasonable solutions to the field equations where the functional relationship can be in a range of forms.

This paper is organized as follows: in the next section we give the background, and prove a number of general results for Weyl-type fields, with or without spatial symmetry.  We show, generalising Majumdar's results for charged dust, that for Majumdar-type fields, the "active gravitational matter density" is numerically equal to the charge density.  We show that for asymptotically flat distributions of charged matter of Majumdar-type, the total mass and total charge, as measured at spatial infinity, are numerically equal.  Finally, we show that any bounded non-singular perfect fluid which is of Majumdar-type consists of dust (i.e. is pressure free).

Sections 3 to 5 looks at the spherically symmetric case.  We present explicit Weyl-type generalisations of the interior Schwarzschild metric, in which the functional relation takes various forms.  As far as the author is aware, these are the first explicit solutions in which the functional relationship deviates from Weyl's original quadratic one. In section 4 a number of these are then investigated as regards their matching conditions with the exterior Reissner-Nordstr\"{o}m solution.  This allows us to identify the constants of integration in terms of the total mass, charge and radius of the sphere.  In section 5 it is also found that many satisfy appropriate regularity and energy conditions.

\section{Background and General Results}
A space-time $(M,g_{\mu\nu})$ is {\it static} if there exists a timelike hypersurface orthogonal Killing vector $\xi^\mu$ on $M$.  Here, and throughout, Greek letters will take values 1 to 4.  With a suitable choice of co-ordinates $(x^i, t)$ the metric takes the form
\[ds^2=g_{ij}dx^idx^j-e^\omega dt^2,\]
where both the spatial metric $g_{ij}$ and $e^\omega$ only depend on $x^i$, for i,j=1 to 3.

Einstein's field equations say that the Ricci tensor $R_{\mu\nu}$ of $g_{\mu\nu}$ is related to the energy-momentum tensor $T_{\mu\nu}$ by
\[R_{\mu\nu}-{\hf} Rg_{\mu\nu}=-\kappa T_{\mu\nu},\]
where $R$ is the scalar curvature $R=g^{\mu\nu}R_{\mu\nu}$. In the case of charged matter, the energy momentum tensor is made up of two parts, the matter tensor $M_{\mu\nu}$, and the electromagnetic energy tensor $E_{\mu\nu}$:
\[T_{\mu\nu}=M_{\mu\nu}+E_{\mu\nu}.\]
The electromagnetic energy tensor is determined by the (skew-symmetric) electromagnetic tensor $F_{\mu\nu}$
\[{\shf}\kappa E_{\mu\nu}=F_{\mu}^{\2 \alpha}F_{\nu\alpha}-\dv{1}{4} F_{\alpha\beta}F^{\alpha\beta}g_{\mu\nu},\]
which satisfies Maxwell's equations
\[\nabla_\nu F^{\mu\nu}=J^\mu,\]
\bee\label{e:pot}
\nabla_{[\alpha}F_{\mu\nu]}=0.
\ee
Here $\nabla$ is covariant differentiation with respect to the Levi-Civita connection of $g_{\mu\nu}$.  $J^\mu$ is the 4-current density vector:
\[J^\mu={\hf}\kappa\sigma V^\mu,\]
where $V^\mu$ is the 4-velocity and $\sigma$ the proper charge of the charged matter. Equation (\ref{e:pot}) is the integrability condition for the existence of the local 4-potential $\phi_\mu$:
\[F_{\mu\nu}= \nabla_\mu\;\phi_\nu -\nabla_\nu\;\phi_\mu.\]
We will consider only pure electric fields $\phi_\mu=(0,0,0,\phi)$ (the magnetic field is assumed to vanish), noting, however, that a theorem of Bonnor \cite{bonn} allows one to generate pure magnetic fields from pure electric fields.

For static pure electric fields, the non-vanishing components of the electromagnetic energy tensor are 
\[\kappa E_{ij}=-2e^{-\omega}\phi,_{i}\phi,_{j}+e^{-\omega}\phi^{,k}\phi,_{k}g_{ij},\]
\[\kappa E_{44}=\phi^{,k}\phi,_{k}\equiv |d\phi|^2.\]
A comma here indicates differentiation, and $\phi$ is the electric potential:
\[F_{i4}=\phi,_{i}\qquad F_{ij}=0.\]
Note that $\phi$ is only defined up to an additive constant $\phi\rightarrow\phi+C$. 

In addition, the 44 component of the Ricci tensor is determined by $\omega$
\[R_{44}=-{\hf} e^\omega(\triangle\omega+{\hf}|d\omega|^2),\]
where $\triangle$ is  the covariant laplacian $\triangle\omega=g^{ij}\omega_{|ij}$, a stroke representing covariant differentiation with respect to the Levi-Civita connection of the spatial metric $g_{ij}$.

  Maxwell's equations reduce to
\bee
\triangle\phi={\hf}\omega^{,i}\phi,_{i}+{\hf}\kappa \sigma e^{{\shf}\omega}.\label{e:max}
\ee
The Einstein equations are
\beer 
R_{ij}&=&2e^{-\omega}\phi,_{i}\phi,_{j}-\kappa M_{ij} +({\hf}\kappa M^\alpha_\alpha -e^{-\omega}|d\phi|^2)g_{ij}\label{e:ein1}\\
-{\hf} e^\omega(\triangle\omega+{\hf}|d\omega|^2)&=&-{\hf}\kappa(M_k^k-M_4^4)e^\omega- |d\phi|^2\label{e:ein2}\\
R&=&\kappa M^\alpha_\alpha.\label{e:ein3}
\eer
Equations (\ref{e:max}), (\ref{e:ein1}), (\ref{e:ein2}) and (\ref{e:ein3}) are the equations governing the static electrogravitational field.  A {\it Weyl-type} field is an electric field satisfying the above equations where $g_{44}$ and $\phi$ are functionally related: $\omega\equiv\omega(\phi)$.

\begin{thm}
For any Weyl-type field the following relationship holds between the gravitational matter density, the electric field density and the charge density:
\bee\label{e:agam}
M^k_k-M^4_4+\left[\frac{d^2e^\omega}{d\phi^2} -2\right]E^4_4=\frac{d\;e^{{\shf}\omega}}{d\phi}\sigma.
\ee
\end{thm}
\noindent{\bf Proof:}
 
The result comes from comparing the 44 field equation (\ref{e:ein2}) with Maxwell's equation (\ref{e:max}).  Assuming that $\omega=\omega(\phi)$ we have that
\[\omega,_{i}=\frac{d\omega}{d\phi}\;\phi,_{i} \qquad \qquad |d\omega|^2=\left(\frac{d\omega}{d\phi}\right)^2\;|d\phi|^2,\]
\[\omega^{,k}\phi,_{k}=\frac{d\omega}{d\phi}\;|d\phi|^2 \qquad\qquad \triangle\omega=\frac{d\omega}{d\phi}\;\triangle\phi+\frac{d^2\omega}{d\phi^2}|d\phi|^2.\]
Substituting these in (\ref{e:ein2}) we get
\[-{\hf} e^\omega\left[\frac{d\omega}{d\phi}\;\triangle\phi +\frac{d^2\omega}{d\phi^2}\;|d\phi|^2+{\hf} \left(\frac{d\omega}{d\phi}\right)^2\;|d\phi|^2\right]=-{\hf}\kappa(M^k_k-M^4_4)e^\omega-|d\phi|^2,\]
and then using Maxwell's equation  (\ref{e:max})
\[-{\hf} e^\omega\left[{\hf}\kappa\frac{d\omega}{d\phi}\;\sigma e^{{\shf}\omega} +\left(\frac{d^2\omega}{d\phi^2} +\left(\frac{d\omega}{d\phi}\right)^2\right)\;|d\phi|^2\right]=-{\hf}\kappa(M^k_k-M^4_4)e^\omega-|d\phi|^2.\]
Rearranging this equation gives the result. $\Box$
\vspace{.2in}

This theorem generalises Majumdar's result  \cite{maj}:

\begin{cor}
For a pure electric field ($M_{\mu\nu}=0$ and $\sigma=0$), the only possible form of a functional relationship $\omega=\omega(\phi)$ is
\[ e^\omega=A+B\phi+\phi^2 \qquad A,B\quad constant.\]
\end{cor}

We also have the following two corollaries:

\begin{cor}
For any Weyl-type field satisfying the relation $ e^\omega=A+B\phi+\phi^2$
\[\left[\frac{M_k^k-M^4_4}{\sigma}\right]^2=\lambda e^{-\omega}+1,\]
where $\lambda=\left(\frac{B}{2}\right)^2-A$.
\end{cor}

\begin{cor}\label{c:3}
For a Majumdar-type field (i.e. one where $ e^\omega=(\frac{B}{2}\pm\phi)^2$)
the following holds:
\[M_k^k-M_4^4=\pm\sigma.\]
\end{cor}
Thus the active gravitational matter density and the charge density of a Majumdar-type field are numerically equal.  A further feature of this balancing in Majumdar-type fields is that the total mass and total charge are equal:

The total mass of a static asymptotically flat space-time $(M,g_{\mu\nu})$ is given by (Wald \cite{wald})
\[m=\lim_{R\rightarrow\infty}\frac{1}{4\pi}\int_{S_R}\frac{\xi^\mu}{(-\xi_\al\xi^\al)^{\shf}}\nabla_{\mu}\xi_{\nu}N^\nu\;dA,\]
where $S_R$ is a sphere of radius R in the hypersurface orthogonal to the Killing vector $\xi^\mu$, $N^\mu$ is the outward pointing normal to $S_R$ which is orthogonal to $\xi^\mu$, and integration is with respect to the volume element induced on $S_R$ by $g_{ij}$.

On the other hand the total charge is given by (Wald \cite{wald} Synge \cite{syng})
\[q=\lim_{R\rightarrow\infty}\frac{1}{4\pi}\int_{S_R}F_{\mu\nu}N^\mu t^\nu \;dA,\]
where $t^\mu$ is the unit normal to the volume enclosed by $S_R$.

\begin{thm}
For an asymptotically flat Majumdar-type field $m=|q|$.
\end{thm}

\noindent{\bf Proof:}
 
The choice of co-ordinates we have made implies that the Killing vector is given by
\[\xi^\mu=\delta_4^\mu \qquad \xi_\mu=-e^\omega\delta^4_\mu,\]
so that
\[\xi^\mu\nabla_\mu\;\xi_{\nu}={\hf} e^\omega\omega,_{\nu}={\hf} e^\omega\; \frac{d\omega}{d\phi}\; \phi,_{\nu},\]
and the total mass is given by
\[
m=\lim_{R\rightarrow\infty} \frac{1}{8\pi}\int_{S_R}e^{{\shf}\omega}\;\frac{d\omega}{d\phi}\; \phi,_kN^k dA
.\]
On the other hand the total charge is
\[
q= \lim_{R\rightarrow\infty }\frac{1}{4\pi}\int_{S_R }e^{-{\shf}\omega}\phi,_kN^k dA
.\]
For a source satisfying Weyl's condition
\[ e^{\omega} = A+B\phi+\phi^2 ,\]
for which the surface of the body is an equipotential surface, Majumdar \cite{maj} has 
shown that this relationship must continue to hold outside the source 
(see also Gatreau and Hoffman \cite{gah}).  In any case, we will assume that Weyl's condition holds throughout the space-time.  Then
\beer
 m&=&\lim_{R\rightarrow\infty} \frac{1}{8\pi}\int_{S_R}\frac{B+2\phi}{\sqrt{A+B\phi+\phi^2}}\phi,_kN^k dA\nn\\
&=&\lim_{R\rightarrow\infty} \frac{1}{8\pi}\int_{S_R}(B+2\phi)e^{-{\shf}\omega}\phi,_kN^k dA\nn
.\eer
We assume that the space-time is asymptotically flat, so that $\phi=O(\dv{1}{r})$ and $A=1$.  Then
\beer
m&= &\lim_{R\rightarrow\infty} \frac{B}{8\pi}\int_{S_R}e^{-{\shf}\omega}\phi,_kN^k dA\nn\\
&=&\frac{Bq}{2}\nn
.\eer
For asymptotically flat Majumdar-type fields $B=\pm 2$ and so $m=|q|$ as claimed. $\Box$

\vspace{.2in}

Finally we look at the compatibility of Weyl's condition with a perfect fluid matter source.  Thus, we assume that the matter tensor is of the form
\[
M^\mu_\nu= (p+\rho)V^\mu V_\nu + p\delta^\mu_\nu
,\]
where $V^\mu$ is the 4-velocity of the fluid and $p$, $\rho$ are the matter 
pressure and density, respectively. We will show that for a static field, 
Majumdar's relation cannot hold for any physically realistic perfect fluid unless the 
matter pressure vanishes. This is one of the consequences of the following:

\begin{thm}
For any static electrogravitational field in which the matter tensor is a 
perfect fluid, the following hold:
\begin{description}
\item{(i)}
If any two of the surfaces of constant $g_{44}$, $\phi$ or $p$ coincide, then 
the third also coincides.
\item{(ii)}
If the fluid is of Weyl-type, then either the pressure gradient 
vanishes at the surface of the fluid or the surface is an equipotential 
surface (i.e. $\phi$ and $e^\omega$ are constant on the surface). 
\item{(iii)}
If the fluid satisfies Majumdar's relation (\ref{e:majrel}), then the pressure satisfies 
\[ p=Ce^{\omega} \qquad C\; constant. ,\]
and the spatial slice is conformal to a space of constant curvature.
\end{description}
\end{thm}

\noindent{\bf Proof:}
 
\noindent {\bf (i)}: To prove this we consider the conservation equation
\[ \nabla_\mu T^\mu_{\nu} =0.\]
For static electrogravitational fields the conservation equation $\nu=i$ reads
\bee\label{e:cons}
p,_{i}+{\hf} (p+\rho)\omega,_{i} = \sigma e^{-{\shf}\omega}\phi,_{i}
.\ee
 From this we see that if any two of $p,_{i}$, $\omega,_{i}$ and 
$\phi,_{i}$ are coincident, then all three are. This proves (i).

\vspace{.1in}

\noindent {\bf (ii)}: We prove this by using the Synge-O'Brien junction conditions 
\cite{sob}. These state that, if $\Sigma$ is the spatial surface formed by the 
boundary of the body with normal $\eta_{i}$, then
\[G_{ij}\eta^{i}\eta^{j} \mbox{ is continuous across } \Sigma.\]
Now, since the electromagnetic field is continuous across $\Sigma$ (see e.g.
Ehlers \cite{ehl})
\[M_{ij}\eta^{i}\eta^{j} \mbox{ is continuous across } \Sigma,\]
and hence, for a perfect fluid we find that
\bee\label{e:pzero}
p(\Sigma)=0
,\ee
i.e. the surface of the body is a surface of constant (zero) pressure.

 Therefore if $g_{44}$ and $\phi$ are functionally related and the pressure 
gradient does not vanish on $\Sigma$, then, by part (i), $\Sigma$ is an equipotential
 surface, as required.
\vspace{.1in}

\noindent {\bf (iii)}: Finally, note that, for a perfect fluid satisfying Majumdar's condition, Corollary \ref{c:3} tells us that
\[3p+\rho=\pm\sigma.\]
Plugging this and Majumdar's relation into the conservation equation (\ref{e:cons}), we find that
\[ p,_{i}=p\omega,_{i} ,\]
which integrates up to
\bee\label{e:press}
p=Ce^{\omega}
,\ee
for some constant C.
Thus we find that
\[M_{ij}=Ce^{\omega}g_{ij}.\]
Now, inserting this in the field equations (\ref{e:ein1}) and making a conformal change of metric $g_{ij}=e^{-\omega}\overline{g}_{ij}$, we find that the Ricci tensor of $\overline{g}_{ij}$ is 
\[^3\overline{R}_{ij}=2C\overline{g}_{ij} ,\]
and this is the defining equation for a 3-dimensional space of constant curvature. $\Box$

\vspace{.2in}

From this we can draw the conclusion that any bounded non-singular perfect fluid satisfying Majumdar's relation must be dust (i.e. pressure-free). This follows
 from the fact that if $C \not=0$ in (\ref{e:press}) then the boundary condition (\ref{e:pzero}) will mean that at the surface $\Sigma$ we must have
\[ e^{\omega(\Sigma)}=0 ,\]
which is impossible, unless the space-time is singular at the boundary.  Thus the matter pressure must vanish throughout.  In the next section we look at a number of spherically symmetric perfect fluid solutions with various functional relationships between $e^\omega$ and $\phi$.  There we will see an example of this phenomenon of the incompatibility of Majumdar's relation with a bounded perfect fluid, unless the matter pressure vanishes.

\section{Spherically Symmetric Solutions}

In this section we illustrate how Weyl's assumption
\[
g_{44} = g_{44}(\phi)
,\]
can be used to obtain exact solutions of the Einstein-Maxwell field 
equations in the case where neither the matter density nor the matter pressure 
vanish. In order to do this, we restrict our attention to spherical symmetry and perfect fluidity (i.e. isotropic matter pressure). 

The case of static spherically symmetric perfect fluids has been investigated 
by numerous authors and many exact solutions have been extracted (see for 
example Buchdahl \cite{buch}, Buchdahl and Land \cite{bal}, Klein \cite{kln}, Tolman \cite{tol}, Whittaker \cite{whit}). Perhaps the best known is the interior solution of Schwarzschild \cite{schw}, in which the matter density is taken to be constant. This neutral 
solution has been generalised to include electromagnetic charge by Bohra and Mehra \cite{bam}, Cooperstock and De La Cruz \cite{cdl}, Florides \cite{flor2}, Mehra \cite{meh1} \cite{meh2} and Patel and Tikekar \cite{pat}. We will also present a number of charged versions of this familiar solution in what follows, one class of which 
contains the solution of Cooperstock and De La Cruz as a special case.
However, the motivation for considering such solutions here is not just to 
add to this list, but to show that Weyl's condition can lead to exact 
solutions when the field is neither pure electric nor charged dust.  Moreover, we will show that the functional relation is not constrained to be Weyl's original quadratic one (\ref{e:weylrel}).

Our first task will be to formulate the static Einstein-Maxwell field 
equations when the spatial metric exhibits spherical symmetry. The metric for 
such a distribution can be written in local co-ordinates ($r$, $\theta, \varphi$, $t$) as
\[
ds^2=e^{\alpha(r)}dr^2 +r^2(d\theta^2+sin^2\theta d\varphi^2) -e^{\omega(r)}dt^2
.\]

From spherical symmetry and the Maxwell equations we find that 
the only non-vanishing components of $T^\mu_\nu$ are (see Florides \cite{flor1})
\[ \kappa T^1_1 =\kappa p -\frac{Q^2}{r^4} \qquad\qquad \kappa T^4_4=-\kappa\rho-\frac{Q^2}{r^4}\]
\[\kappa T^2_2=\kappa T^3_3=\kappa M^2_2+\frac{Q^2}{r^4},\]
where $p$ and $\rho$ are the radial matter pressure and density, and $Q$ is the 
total charge inside a radius $r$, i.e.
\beer
p(r) &\equiv & M^1_1 \nn\\
\rho(r) &\equiv & -M^4_4 \nn \\
Q(r) &\equiv & {\dv{\kappa}{2}}\int_0^r\sigma(r)r^2e^{{\shf}\alpha}dr \nn\\
 &=& r^2\phi'e^{-{\shf}(\alpha+\omega)} \nn
.\eer
Now, the field equations 
\[ \kappa T^\mu_\nu = -G^\mu_\nu,\]
when we insert the spherical metric form and the energy momentum tensor, read
\beer
\kappa p-\frac{Q^2}{r^4} &=& -\frac{1}{r^2} +\frac{e^{-\alpha}}{r^2}(1+r\omega') \nn\\
   \noalign{\vskip 6pt}
\kappa M^2_2+\frac{Q^2}{r^4} &=& e^{-\alpha}\left({\hf}\omega''-{\dv{1}{4}}\alpha'\omega'+
  {\dv{1}{4}}\omega'^2 +\frac{\omega'-\alpha'}{2r}\right) \nn\\
   \noalign{\vskip 6pt}
\kappa \rho +\frac{Q^2}{r^4} &=& \frac{1}{r^2}-\frac{e^{-\alpha}}{r^2}(1-r\alpha')\nn
,\eer
where a prime denotes differentiation with respect to. $r$.

The above equations constitute the most general system of 
equations for any static charged sphere in general relativity. However, we still have six 
unknowns ($p$, $\rho$, $M^2_2$, $Q$, $\alpha$, $\omega$) and only three equations.  To make the system determinate we will first assume that the fluid is perfect ($M_1^1=M_2^2=M_3^3$).  In this case the field equations become
\beer
\kappa p-\frac{Q^2}{r^4} &=& -\frac{1}{r^2} +\frac{e^{-\alpha}}{r^2}(1+r\omega') \label{e:feq1} \\
   \noalign{\vskip 6pt}
\kappa p+\frac{Q^2}{r^4} &=& e^{-\alpha}\left({\hf}\omega''-{\dv{1}{4}}\alpha'\omega'+{\dv{1}{4}}\omega'^2 +\frac{\omega'-\alpha'}{2r}\right) \label{e:feq2}  \\
   \noalign{\vskip 6pt}
\kappa \rho +\frac{Q^2}{r^4} &=& \frac{1}{r^2}-\frac{e^{-\alpha}}{r^2}(1-r\alpha') \label{e:feq3}
.\eer

Following Tolman \cite{tol} we note that subtracting (\ref{e:feq2}) from (\ref{e:feq1}) we get
\bee\label{e:feq4}
\frac{d}{dr}\left(\frac{e^{-\alpha}-1}{r^2}\right) +\frac{d}{d r}
  \left(\frac{\omega'e^{-\alpha}}{2r}\right) +e^{-(\alpha+\omega)}\frac{d}{d r}\left(\frac
   {\omega'e^{\omega}}{2r}\right) =\frac{4Q^2}{r^5}
.\ee

Our plan of attack for solving the field equations (\ref{e:feq1}) to (\ref{e:feq3}) will be to 
introduce two conditions which allow us to integrate (\ref{e:feq4}) fully for 
$\omega$ and $\alpha$ as explicit functions of $r$. Then we substitute these in 
(\ref{e:feq1}) to (\ref{e:feq3}) and determine $p(r)$, $Q(r)$ and $\rho(r)$.

The first condition we impose is the Schwarzschild condition
\bee\label{e:schwarz}
-T^4_4=\kappa\rho +\frac{Q^2}{r^4}= \frac{3}{R^2} = constant
.\ee
Equation (\ref{e:feq3}) implies that
\bee\label{e:aldef}
e^{-\alpha}=1-\frac{r^2}{R^2}
.\ee
Plugging this in (\ref{e:feq4}) we find that
\[\frac{d}{dr}\left(\frac{\omega'e^{-\alpha}}{2r}\right)+e^{-(\alpha+\omega)}\frac{d}{dr}
  \left(\frac{\omega'e^{\omega}}{2r}\right)=\frac{4Q^2}{r^5} ,\]
or multiplying across by \(\frac{2\omega'e^{\omega}}{r}\)
\[\frac{d}{dr}\left[\frac{\omega'^2e^{\omega-\alpha}}{r^2}\right]=\frac{8Q^2\omega'e^{\omega}}
   {r^6} ,\]
\beer
\Rightarrow \frac{d}{dr}\left[\ln \left(\frac{\omega'e^{{\shf}(\omega-\alpha)}}{r}\right)
   \right]&=&\frac{4Q^2e^{\alpha}}{r^4\omega'}\nn\\
&=&\frac{4(\phi')^2}{(e^\omega)'}\nn
.\eer
Now we introduce Weyl's condition $\omega=\omega(\phi)$, or, more conveniently, $\phi=\phi(e^\omega)$.  Then we have that
\[ \frac{d}{dr}\left[\ln \left(\frac{\omega'e^{{\shf}(\omega-\alpha)}}{r}\right)
   \right]=4\left(\frac{d\phi}{d\;e^\omega}\right)^2(e^\omega)',\]
or integrating
\bee\label{e:ddef}
\ln \left(\frac{\omega'e^{{\shf}(\omega-\alpha)}}{r}\right)
   =4\int\left(\frac{d\phi}{d\;e^\omega}\right)^2d(e^\omega)+\ln 2D
,\ee
for some positive constant $D$.  Now, define 
\bee\label{e:func}
f(e^\omega)\equiv \exp\left[-4\int\left(\frac{d\phi}{d\;e^\omega}\right)^2d(e^\omega)\right]
.\ee
Then a further integration yields
\beer
\int f(e^\omega)d(e^{{\shf}\omega})&=&D\int\frac{r}{(1-\frac{r^2}{R^2})^{{\shf}}}dr\nn\\
&=&F-DR^2\left(1-\frac{r^2}{R^2}\right)^{{\shf}}\label{e:solv}
,\eer
where $F$ is another constant of integration.

Our strategy to solve the equations will be to choose a functional relation $\phi=\phi(e^\omega)$, integrate (\ref{e:func}) to get $f(e^\omega)$ and then integrate (\ref{e:solv}) and invert to find $e^\omega$ as a function of $r$.  We will investigate two types of functional relation and their corresponding solutions, which we refer to as {\it Class I} and {\it Class II solutions}.

\subsection{Class I Solutions}

First we consider fields in which
\bee\label{e:class1}
Ae^{\omega}=B+(C+\phi)^2
,\ee
where $A$, $B$ and $C$ are constants.
For $A=1$ this is precisely Weyl's original condition (\ref{e:weylrel}), and if, in addition, $B=0$, this 
reduces to Majumdar's condition (\ref{e:majrel}). Indeed the appearance of $A$ in this equation 
may at first seem superfluous, at least for $A>0$, as we have the freedom to 
rescale our time co-ordinate
\[
 t \rightarrow t'= A^{{\shf}}t 
,\]
so that
\[ Ae^{\omega} \rightarrow e^{\omega} .\]
However, if we initiate this co-ordinate change, we must also change the 
electromagnetic tensor $F_{\mu\nu}$, and it is easy to check that this change of 
co-ordinates induces a rescaling of $\phi$
\[\phi \rightarrow \frac{\phi}{A^{{\shf}}} .\]
Hence, a change of co-ordinates of this type does not alter the form of 
the functional relationship, since it merely induces a rescaling of the constants $B$ and $C$.
We shall see that in order to integrate the field equation with $B\ne 0$, we will allow $A$ to only take on the values 1, ${\hf}, \dv{3}{2}$ and that 
each of these leads to significantly different solutions.  For $A\ne 1$ we are considering for the first time (as far as the author is aware) a functional relation different from the original one introduced by Weyl.

Taking this relation then and computing $f(e^\omega)$ from (\ref{e:func}) we have that
\[f(e^\omega)=(Ae^\omega-B)^{-A},\]
and then (\ref{e:solv}) tells us that
\[\int \frac{d(e^{{\shf}\omega})}{(Ae^\omega-B)^A}=F-DR^2\left(1-\frac{r^2}{R^2}\right)^{{\shf}}.\]
For $A=0$ this gives the standard interior Schwarzschild solution.
For $B=0$ this can be integrated up to give $e^{\omega(r)}$ for any value of $A$ (although we require that $A\ge0$ for the functional relationship to be real).  For $B\not= 0$ we can integrate this completely if
\[ A = \pm n, \pm (n+{\hf}) \qquad \qquad n \in \mbox{\boldmath $N$} .\]
However, in general, we then get an implicit equation for $e^{\omega}$ in terms 
of $r$. The only values of $A$ that lead to an explicit expression for $e^\omega$ for $B\not= 0$ are 
1, ${\hf}, \dv{3}{2}$.

Thus the logic of the field equations forces us to consider the cases $B=0$ ({\it Class Ia})
and $B\not= 0$ ({\it Class Ib}) separately.  We now list the metric components, electric charge, density and pressure of each solution.  Recall that for each one
\[e^{-\alpha}=1-\frac{r^2}{R^2}.\]
For brevity, we have defined
\[
\Phi(r)\equiv F-DR^2\left(1-\frac{r^2}{R^2}\right)^{{\shf}}
.\]

\noindent \ul{{\large\bf Class Ia}}

\beer
A\ne{\hf}\qquad& & \nn\\
e^{\omega}&=& \left[(1-2A)A^A\Phi\right]^{\frac{2}{1-2A}} \nn\\
   \noalign{\vskip 6pt}
Q(r)&=&\pm\frac{\sqrt{A}\; Dr^3}{(1-2A)\Phi} \nn\\
   \noalign{\vskip 6pt}
\kappa\rho(r)&=&\frac{3}{R^2}-\frac{AD^2r^2}{(1-2A)^2\Phi^2} \nn\\
\kappa p(r)&=&-\frac{1}{R^2}+\frac{AD^2r^2}{(1-2A)^2\Phi^2}+
   \frac{2D(1-\frac{r^2}{R^2})^{{\shf}}}{(1-2A)\Phi} \nn\\
   \noalign{\vskip 6pt}
   \noalign{\vskip 6pt}
A={\hf}\qquad& & \nn\\
\omega&=& \sqrt{2}\;\Phi(r)\nn\\
   \noalign{\vskip 6pt}
Q(r)&=&\pm\frac{Dr^3}{2} \nn\\
   \noalign{\vskip 6pt}
\kappa\rho(r)&=&\frac{3}{R^2}-\frac{D^2r^2}{4}\nn\\
   \noalign{\vskip 6pt}
\kappa p(r)&=&-\frac{1}{R^2}+\frac{D^2r^2}{4}+\sqrt{2}\;D\left(1-\frac
   {r^2}{R^2}\right)^{{\shf}}\nn
.\eer

\noindent \ul{{\large\bf Class Ib}}

\begin{description}
\item{\ul{$A={\hf}$}:}
\beer
e^{\omega}&=&\left\lbrace\begin{array}{lr}
  2B\cosh^2\left[\frac{1}{\sqrt{2}} \Phi(r) \right] & for \qquad B>0\\
  \noalign{\vskip 6pt}
  -2B\sinh^2\left[\frac{1}{\sqrt{2}} \Phi(r) \right] & for \qquad B<0
\end{array}\right.\nn\\
Q(r)&=&\pm\frac{Dr^3}{2} \nn\\
   \noalign{\vskip 6pt}
\kappa\rho(r)&=&\frac{3}{R^2}-\frac{D^2r^2}{4} \nn\\
   \noalign{\vskip 6pt}
\kappa p(r)&=&  \frac{1}{R^2}+\frac{D^2r^2}{4}+\left(1-\frac{r^2}{R^2}\right)
^{{\shf}}\!\!.\left\{\begin{array}{lr}
  \sqrt{2}D\tanh\left[\frac{\Phi}{\sqrt{2}}\right] & for \qquad B>0 \nn\\
  \noalign{\vskip 6pt}
  \sqrt{2}D\coth\left[\frac{\Phi}{\sqrt{2}}\right] & for \qquad B<0
\end{array}\right.\nn
\eer
\vspace{.2in}

\item{\ul{A=1}:}
\beer
e^{\omega}&=&\left\lbrace\begin{array}{lr}
  B\coth^2\left[\sqrt{B} \Phi(r) \right] & for \qquad B>0 \nn\\
  \noalign{\vskip 6pt}
  -B\tan^2\left[\sqrt{-B} \Phi(r) \right] & for \qquad B<0
\end{array}\right.\nn\\
Q(r)&=& \left\lbrace\begin{array}{lr}
 \pm\sqrt{B}Dr^3cosech[\sqrt{B}\:\Phi] & for \qquad B>0 \nn\\
  \noalign{\vskip 6pt}
 \pm\sqrt{-B}Dr^3\sec[\sqrt{-B}\:\Phi] & for \qquad B<0
\end{array}\right.\nn\\
\noalign{\vskip 6pt}
\kappa\rho(r)&=&\left\lbrace\begin{array}{lr}
  \frac{3}{R^2}-BD^2r^2cosech^2[\sqrt{B}\:\Phi] & for \qquad B>0 \nn\\
  \noalign{\vskip 6pt}
  \frac{3}{R^2}+BD^2r^2\sec^2[\sqrt{-B}\:\Phi] & for \qquad B<0
\end{array}\right.\nn\\
\noalign{\vskip 6pt}
\kappa p(r)&=& \left\lbrace\begin{array}{lr}
 -\frac{1}{R^2}+BD^2r^2cosech^2[\sqrt{B}\:\Phi] & for \quad B>0\nn\\
  \qquad+2\sqrt{B}Dsech[\sqrt{B}\:\Phi]cosech[\sqrt{B}\:\Phi]\left(1-\frac{r^2}{R^2}
   \right)  ^{{\shf}} & \nn\\
\noalign{\vskip 6pt}
  -\frac{1}{R^2}-BD^2r^2\sec^2[\sqrt{-B}\:\Phi] & for \quad B<0\nn\\
  \qquad+2\sqrt{-B}D\sec[\sqrt{-B}\:\Phi]\csc[\sqrt{-B}\:\Phi]\left(1-\frac{r^2}{R^2}
   \right)  ^{{\shf}}&\nn
\end{array}\right.\nn
\eer

\vspace{.2in}

\item{\ul{A=$\dv{3}{2}$}:}
\beer
e^{\omega}&=&\frac{2B^3\left[F-DR^2\left(1-\frac{r^2}{R^2}\right)^{{\shf}}\right]^2}
{3B^2\left[F-DR^2\left(1-\frac{r^2}{R^2}\right)^{{\shf}}\right]^2-2}\nn\\
Q(r)&=&\pm\frac{3DBr^3}{3B^2\Phi^2-2} \nn\\
   \noalign{\vskip 6pt}
\kappa\rho(r)&=&\frac{3}{R^2}- \frac{9B^2D^2r^2}{(3B^2\Phi^2-2)^2} \nn\\
   \noalign{\vskip 6pt}
\kappa p(r)&=&-\frac{1}{R^2}+ \frac{9B^2D^2r^2}{(3B^2\Phi^2-2)^2} -
  \frac{4D\left(1-\frac{r^2}{R^2}\right)^{{\shf}}}{\left(3B^2\Phi^2-2\right)
  \Phi} \nn
.\eer
\end{description}

\vspace{.2in}

In sections 4 and 5 we will look at the features of some of these solutions more closely.

\subsection{Class II Solutions}

We now consider fields in which
\[
\dv{\sqrt{2}A}{3}e^{\sdv{3}{4}\omega}=C+\phi 
,\]
where $A$ and $C$ are constants.  For $A=0$ we should get the interior Schwarzschild solution.

Taking this relation and computing $f(e^\omega)$ from (\ref{e:func}) we have that
\[f(e^\omega)=\exp(-A^2e^{{\shf}\omega}),\]
and then (\ref{e:solv}) tells us that
\[-\frac{1}{A^2}\exp(-A^2e^{{\shf}\omega}) = F-DR^2\left(1-\frac{r^2}{R^2}\right)^{{\shf}},\]
or
\[
e^\omega=\frac{1}{A^4}\left[\ln \left(-A^2\left(F-DR^2 (1-\frac{r^2}{R^2})^{{\shf}}\right)\right)\right]^2
.\]
The electric charge, density and pressure of this solution turn out to be:

\beer
Q(r)&=&\pm \frac{Dr^3}{\sqrt{2}\Phi} \left[-\ln (-A^2\Phi)\right]^{-{\shf}} \nn\\
   \noalign{\vskip 6pt}
\kappa\rho(r)&=&\frac{3}{R^2}-  \frac{D^2r^2}{2\Phi^2}\left[-\ln (-A^2\Phi)\right]^{-1} \nn\\
\noalign{\vskip 6pt}
\kappa p(r)&=& -\frac{1}{R^2}+\frac{D}{\Phi^2} \left[-\ln (-A^2\Phi)\right]^{-1}\left[{\hf} Dr^2-2\left(1-\frac{r^2}{R^2}\right)^{{\shf}}\Phi\right] \nn
.\eer
Note that we must have $0<-A^2\Phi(r)<1$ to have real a solution.

\section{Physical Significance of the Constants}

In this section we relate the constants which appear in the previous solutions with the more physically relevant constants of total mass, total charge and radius of the body. To do this we consider the boundary conditions required of 
the solutions for them to match with the unique exterior solution. Before looking at some of these solutions in detail, we first draw a number of conclusions which come from their general form.

Outside of the matter distribution we must have the 
Reissner-Nordstr\"{o}m metric
\[
e^{-\alpha}_{rn}=e^{\omega}_{rn}=1-\frac{2m}{r}+\frac{q^2}{r^2}
,\]
If we let $r=a$ be the boundary of the distribution, then $m$, the 
total mass, and $q$, the total charge of the body are given by (Florides \cite{flor1})
\beer
m&=&{\hf}\int_0^a\left(\kappa\rho+\frac{Q^2}{r^4}\right) r^2dr
  +\frac{q^2}{2a} \label{e:mass}\\
q&=&Q(a)\label{e:chrg}
.\eer 

Our aim will be to relate the arbitrary constants which arose in the integration of the field equations to the physically significant constants $m$, $q$ and $a$. 
We do this by using the continuity conditions which we expect 
the metric to exhibit across the surface $r=a$.

In particular, we demand that $e^{\alpha}$, $e^{\omega}$ and $\omega'$ all 
be continuous across $r=a$. (We note that as a consequence of the 
continuity of $\omega'$ the pressure $p$ vanishes at $r=a$ by (\ref{e:feq1}) - as expected).  Hence 3 constants in each solution will be determined by the 
boundary conditions at $r=a$. Furthermore a 4th constant will also be expressible in terms of $m$, $q$ and $a$ by evaluating (\ref{e:chrg}) at $r=a$.  Therefore, all of the constants appearing in the solutions presented (aside from $C$, the phase of the electric field, which we will ignore from here on) can be written in terms of the total mass, total charge and radius of the source. The actual relationships between these constants will obviously depend on the particular solution under consideration.  However all the solutions have in common the Schwarzschild condition
\[
\kappa\rho +\frac{Q^2}{r^4}= \frac{3}{R^2} = constant
,\]
which when put in (\ref{e:mass}) gives
 \bee\label{e:rconst}
\frac{1}{R^2}=\frac{2}{a^3}\left( m-\frac{q^2}{2a} \right)
.\ee
This is equivalent to the continuity of $e^\alpha$ across the boundary $r=a$.

As a final comment we note that for the charge density to be non-singular at 
the origin, $r=0$, we must have
\[ Q(r) \sim r^3 \qquad as \qquad r \rightarrow 0 ,\]
and this condition is satisfied by all of the new fields presented.

\subsection{Class I Solutions}

All of these solutions contain the constants $D$, $F$, $R$ and 
either $A$ or $B$.  The constant $R$ is given in terms of $m$, $q$ and $a$ by (\ref{e:rconst}), and $A$ is positive for each of the solutions.

We determine $A$ and $B$ in terms of the total mass, total charge and radius as follows. Recall that the charge inside a radius $r$ is
\[Q(r)= r^2\phi'e^{-{\shf}(\alpha+\omega)},\]
and given the functional relationship
\[Ae^\omega=B+(C+\phi)^2,\]
this becomes
\[ Q(r)= \pm\frac{Ar^2\omega'e^{-{\shf}(\alpha-\omega)}}{2\sqrt{Ae^\omega-B}}.\]
Now using (\ref{e:chrg}) and the Reissner-Nordstr\"{o}m solution we get that
\[q^2=\frac{A^2\left(m-\frac{q^2}{a}
  \right)^2}{\left(A(1-\frac{2m}{a}+\frac{q^2}{a^2})-B\right)} ,\]
which can be rearranged to read
\bee\label{e:abconst}
B=A\left[1-\left(\frac{m}{q}\right)^2+(1-A)\left(\frac{m}{q}-\frac{q}{a}
  \right)^2\right]
.\ee
For the Class Ia solutions $B=0$ and this gives $A$ in terms of $m$, $q$ and $a$.  In this case we note that
\[A\ge1 \qquad \Leftrightarrow \qquad m \le |q| .\]
Moreover, this relation means that
\[A\le1 \qquad \Rightarrow \qquad 1\leq\left(\frac{m}{q}\right)^2\le\frac{1}{A},\]
\[A\ge1 \qquad \Rightarrow \qquad \frac{1}{A} \leq\left(\frac{m}{q}\right)^2\le1.\]
Thus, $A$ determines the balancing of the total mass and total charge.

For the Class Ib solutions $B\not=0$ and A is 1 or ${\hf}$ or $\dv{3}{2}$.  Thus, equation (\ref{e:abconst}) this gives $B$ in terms of $m$, $q$ and $a$.  In this case we have that
\[ A\geq1 \quad and \quad  B\ge0 \qquad \Rightarrow \qquad m\le |q| ,\]
\[
A\le1 \quad and \quad B\le0 \qquad \Rightarrow \qquad m\ge |q|
.\]

Turning now to the constant $D$, we note that putting our functional relationship in (\ref{e:ddef})  gives
\bee\label{e:omprime1}
\frac{\omega'e^{{\shf}(\omega-\alpha)}}{r}=2D(Ae^\omega-B)^A
.\ee
Since we require $\omega'$ to be continuous across $r=a$, we use the Reissner-Nordstr\"{o}m solution and (\ref{e:abconst}) to get that
\beer
D&=&\frac{1}{a^3}\left(m-\frac{q^2}{a}\right)\left(A(1-\frac{2m}{a}+\frac{q^2}{a^2})-B\right)^{-A}\nn\\
&=&\frac{1}{a^3} \left(m-\frac{q^2}{a}\right)\left|A\left(\frac{m}{q}-\frac{q}{a}\right)\right|^{-2A}\label{e:d1const}
.\eer
This will determine $D$ in terms of the total mass, total charge and radius of the configuration for Class I solutions.

The final constant $F$ will be determined by the requirement that $e^\omega$ be continuous across $r=a$.  This condition will differ for each solution and is rather complex, so we will take three illustrative examples in the Class I solutions.

\subsubsection{Class Ia $A={\hf}$}

Equation (\ref{e:abconst}) yields
\[\left(\frac{m}{q}\right)^2-1={\hf}\left(\frac{m}{q}-\frac{q}{a}\right)^2,\]
and so we have that $m\ge|q|$.  We can solve this for $m$ to get
\[m=-\frac{q^2}{a} \pm\frac{q}{a}\sqrt{2(a^2+q^2)}.\]
Equation (\ref{e:d1const}) reduces to
\[D=\frac{2|q|}{a^3}.\]
The continuity of $e^\omega$ across $r=a$ means $F$ is given by
\[F= \left(\frac{m}{q}-\frac{q}{2a}\right)^{-1}\left(1-\frac{2m}{a}+\frac{q^2}
   {a^2}\right)^{{\hf}}+\sqrt{2}\;\ln \left(1-\frac{2m}{a}+\frac{q^2}{a^2}\right) .\]
As a check, it's not hard to show that $p(a)=0$, as required.

\subsubsection{Class Ia $A=1$}

We note that in this case the functional relationship (\ref{e:class1}) reduces to Majumdar's, and (\ref{e:abconst}) tells us that $m=|q|$, as expected. The other constants work out to be
\[D= \frac{|q|}{a^3}\left(1-\frac{|q|}{a}\right)^{-1} \qquad
R^2=\frac{a^3}{|q|}\left(2-\frac{|q|}{a}\right)^{-1} .\]
To find $F$ note that the pressure is
\[p(r)=\frac{D^2R^4-F^2}{R^2\Phi(r)^2},\]
and so for us to insist that $p(a)=0$ we must have \(F^2=D^2R^4\). Thus the matter pressure vanishes, and we are led to the charged dust solution first 
discovered by Cooperstock and De La Cruz \cite{cdl}.  As this solution has been considered before, we will not discuss this solution further here.

\subsubsection{Class Ib $A=1$ $B<0$}

The constants $B$, $R$ and $D$ are given by
\beer
B= 1-\left(\frac{m}{q}\right)^2 & & \qquad \frac{1}{R^2}= \frac{q}{a^3}
   \left(\frac{2m}{q}-\frac{q}{a}\right) \nn\\
   \noalign{\vskip 6pt}
D &=& \frac{q^2}{a^3}\left(m-\frac{q^2}{a}\right)^{-1}\nn
.\eer
Obviously
\[m\ge|q| \Longleftrightarrow B\le0.\]
We shall restrict our discussion to the undercharged case $B<0$ (i.e. $m>|q|$). In this case the remaining constant $F$ is given by the continuity 
of $\omega'$ across $r=a$:
\vspace{.25in} 
\[\sec\left[\left(\left(\frac{m}{q}\right)^2-1\right)^{{\shf}}\left(F-DR^2
   \left(1-\frac{a^2}{R^2}\right)^{{\shf}}\right)\right]=
  \left(\frac{m}{q}-\frac{q}{a}\right)\left[\left(\frac{m}{q}
   \right)^2-1\right]^{-{\shf}},\]
which we can rearrange to read
\beer
F&=&\left[\left(\frac{m}{q}\right)^2-1\right]^{-{\shf}}\sec^{-1}
   \left[\left(\frac{m}{q}-\frac{q}{a}\right)\left[\left(
   \frac{m}{q}\right)^2-1\right]^{-{\shf}}\right]\nn\\
   \noalign{\vskip 6pt}
& &\qquad+\left(1-\frac{2m}{a}+\frac{q^2}{a^2}\right)^{\shf}\left(\frac{m}{q}
   -\frac{q}{a}\right)^{-1}\left(\frac{2m}{q}-\frac{q}{a}\right)^{-1}\label{e:fdef}
.\eer

\subsection{Class II Solutions}

These solutions contain the constants $A$, $D$, $F$ and $R$.  Equation (\ref{e:rconst}) gives $R$ in terms of $m$, $q$ and $a$.

We determine $A$ in terms of the total mass, total charge and radius as follows. Recall again that the charge inside a radius $r$ is
\[Q(r)= r^2\phi'e^{-{\shf}(\alpha+\omega)},\]
and given the functional relationship
\[\frac{\sqrt{2}}{3}Ae^{\sdv{3}{4}\omega}=C+\phi,\]
this becomes
\[ Q(r)= \frac{Ar^2}{2\sqrt{2}} \omega'e^{-{\shf}\alpha+\sdv{1}{4}\omega} .\]
Now using (\ref{e:chrg}) and the Reissner-Nordstr\"{o}m solution at $r=a$ we get that
\[A=\sqrt{2}\left(\frac{m}{q}-\frac{q}{a}\right)^{-1}\left(1-\frac{2m}{a}+\frac{q^2}{a^2}\right)^{\sdv{1}{4}}
.\]

Turning now to the constant $D$, we note that putting our functional relationship in (\ref{e:ddef})  gives
\bee\label{e:omprime2}
\frac{\omega'e^{{\shf}(\omega-\alpha)}}{r}=2D\exp\left( A^2 e^{{\shf}\omega}\right)
.\ee
Since we require $\omega'$ to be continuous across $r=a$, we again use the Reissner-Nordstr\"{o}m solution to get that
\bee\label{e:d2const}
D= \frac{1}{a^3}\left(m-\frac{q^2}{a}\right)\exp\left[-2\left(\frac{m}{q}-\frac{q}{a}\right)^{-2}\left(1-\frac{2m}{a}+\frac{q^2}{a^2}\right)\right]
.\ee

The final constant $F$ will be determined by the requirement that $e^\omega$ be continuous across $r=a$.   This works out to be
\beer
F&=& {\hf}\left(\frac{m}{q}-\frac{q}{a}\right)e^{{\shf}\om(a)}\left[\left(\frac{m}{q}-\frac{q}{2a}\right)^{-1}- \left(\frac{m}{q}-\frac{q}{a}\right)e^{-\om(a)}\right] \nn\\
&&. \exp\left[-2\left(\frac{m}{q}-\frac{q}{a}\right)^{-2}e^{\om(a)}\right] \nn
,\eer
where $e^{\om(a)}=1-\frac{2m}{a} +\frac{q^2}{a^2}$.

\section{Regularity and Energy Conditions}

In order for a solution of the field equations to be considered physically plausible there are a number of conditions that it must satisfy. In particular, we may require that a solution:

\begin{description}
\item[(a)] be non-singular,
\item[(b)] have no horizons,
\item[(c)] satisfy regularity conditions at the origin,
\item[(d)] satisfy energy conditions.
\end{description}

The first of these simply means that the metric and the energy-momentum tensor are finite throughout.  This is not true for the charged dust solution (Class Ia, $A=1$) first discovered by Cooperstock and De La Cruz \cite{cdl}, and we exclude this solution from our consideration. We will consider the other solutions shortly.

A horizon will exist where $e^\omega=0$. This can happen either outside or inside the matter distribution.  Outside the distribution, horizons can only occur if $m\ge|q|$ and then these can occur at 
\[ r_{\pm}=m\pm\sqrt{m^2-q^2} .\]

However, from equation (\ref{e:rconst}) we see immediately that for all of the solutions 
\[ m>\frac{q^2}{2a}.\]

We also note that our derivation of the solutions has implicitly assumed that $\omega'>0$ ({\it cf.} equation (\ref{e:ddef})). Since we require that $\omega'$ be continuous across $r=a$ we must have
\bee\label{e:masscon}
m>\frac{q^2}{a}
.\ee

We therefore conclude that there will be no Reissner-Nordstr\"{o}m 
repulsion outside of any of the solutions which we have derived. (For a 
full description of this phenomenon see, for example, Papapetrou \cite{pap2}). 

Furthermore, it is not hard to show that
\[r_-\quad\le \quad\frac{q^2}{m}\quad\le \quad r_+ ,\]
and since we cannot have the boundary of the body lying in the non-static 
region between the horizons, we conclude that for all of 
the solutions presented, if $m\geq|q|$ holds, then (by equation (\ref{e:masscon}))
\[a> m+\sqrt{m^2-q^2},\]
i.e. there are no horizons exterior to any of the solutions.  As for the possibility of horizons inside he matter, we will discuss this shortly.

The requirements that the solution be regular at the origin are

\begin{description}
\item[(i)] $\alpha'(0)=0$,
\item[(ii)] $\omega'(0)=0$,
\item[(iii)] $p'(0)=0$.
\end{description}

A straightforward differentiation of equation (\ref{e:aldef}) shows that (i) holds for all interior solutions satisfying the Schwarzschild condition (\ref{e:schwarz}), while equations (\ref{e:omprime1}) and (\ref{e:omprime2}) ensure that regularity condition (ii) holds for all Class I and Class II solutions which are non-singular at the origin.  Below we will consider condition (iii) for the new solutions individually.

Finally there are many different energy conditions that can be imposed on the solutions.  We consider:

\begin{description}
\item[(i)] $-T^4_4\ge 0 $,
\item[(ii)] $T^1_1-T^4_4\ge 0 \qquad $ or $\qquad p+\rho\ge 0\qquad$ {\it the weak energy condition},
\item[(iii)] $T^\alpha_\alpha-T_4^4\ge 0\qquad\qquad\qquad\qquad$ {\it the strong energy condition}.
\end{description}

We could add to these the extra conditions
\begin{description}
\item[(iv)] $\rho\ge 0$,
\item[(v)] $p\ge 0$,
\item[(vi)] $3p+\rho\ge 0$,
\item[(vii)] $p'\ge 0$,
\item[(viii)] $\rho'\le 0$.
\end{description}

Of these, condition (i) is automatically satisfied by all solutions ({\it cf.} equation (\ref{e:schwarz})).  Note also that energy condition (vi) implies the strong energy condition (iii).  To continue our investigation we must look at the explicit solutions separately. We will see that all of the new solutions presented satisfy (i) to (iii), while many of the conditions (iv) to (viii) also hold in some cases.

\subsection{Class Ia $A={\hf}$}

This solution is clearly non-singular everywhere and has no horizons. 

Also, since
\beer
\kappa T^1_1-\kappa T^4_4&=&\frac{2}{R^2}+\sqrt{2}\;D\left(1-\frac{r^2}{R^2}\right)^{{\shf}}\geq0, \nn\\
   \noalign{\vskip 6pt}
\kappa T^{\alpha}_{\alpha}-\kappa T^4_4&=&D^2r^4+3\sqrt{2}\;D\left(1-\frac{r^2}{R^2}\right)^{{\shf}} \geq0, \nn\\
    \noalign{\vskip 6pt}
3\kappa p+\kappa\rho&=&{\hf} D^2r^4+3\sqrt{2}D\left(1-\frac{r^2}{R^2}\right)^{\shf}\geq 0,\nn
\eer
energy conditions (ii), (iii) and (vi) all hold.

Differentiating the expression for the matter pressure gives
\[
\kappa p'=Dr\left[{\shf} D-\frac{\sqrt{2}}{R^2}\left(1-\frac{r^2}{R^2}\right)^{-{\shf}}\right]
.\]
Thus, $p'(0)=0$ and the solution satisfies all of the regularity conditions at $r=0$.  Focusing on the term in square brackets, and using the expressions for $D$ and $R$ in terms of $m$, $q$ and $a$, as well as the fact that $m\geq|q|$ 
\beer
{\shf} D-\frac{\sqrt{2}}{R^2}\left(1-\frac{r^2}{R^2}\right)^{-{\shf}}&\leq& {\shf} D-\frac{\sqrt{2}}{R^2}\nn\\
&=&\frac{|q|}{a^3}-\frac{2\sqrt{2}}{a^3}\left(m-\frac{q^2}{2a}\right)\nn\\
&\leq &\frac{m}{a^3}-\frac{2\sqrt{2}}{a^3}\left(m-\frac{q^2}{2a}\right)\nn\\
&=&-\left(\frac{2\sqrt{2}-1}{a^3}\right)\left(m-\frac{\sqrt{2}}{2\sqrt{2}-1}\frac{q^2}{a}\right)\nn\\
&\leq&0,\nn
\eer
where this last inequality comes from equation (\ref{e:masscon}).  Thus we conclude that $p'\leq 0$ (energy condition (vii)) and since $p(a)=0$, $p\geq 0$ (energy condition (v)).

Finally, differentiating the matter density
\[
\kappa\rho'=-{\hf} D^2 r\leq 0
,\]
satisfying energy condition (viii), and since
\beer
\kappa\rho(a)&=&\frac{3}{R^2}-\frac{D^2a^2}{4}\nn\\
&=&\frac{6}{a^3}\left(m-\frac{2q^2}{3a}\right)\nn\\
&>&0
,\eer
we have $\rho\geq 0$ (condition (iv)).

In conclusion, this solution satisfies all of the regularity conditions and energy conditions (i) to (viii).

\subsection{Class Ib $A=1$ $B<0$}

For this solution to be non-singular and have no horizons, we require that 
\[
0<\left[\left(\frac{m}{q}\right)^2-1\right]^{\shf}\Phi(r)<\frac{\pi}{2}.
\]
Since $\Phi(r)$ is a non-decreasing function of $r$, this is equivalent to
\[
\Phi(0)>0 \qquad and \qquad \Phi(a)<\frac{\pi}{2}.\left[\left(\frac{m}{q}\right)^2- 1\right]^{-{\shf}}
.\]
The second of these follows from the definition of the constant $F$, equation (\ref{e:fdef}).  The first can be seen to hold as follows. 

A short calculation shows that
\bee\label{e:mess}
1-\frac{2m}{a}+\frac{q^2}{a^2}>0 \qquad \Rightarrow \qquad
\left[\left(\frac{m}{q}\right)^2-1\right]^{-{\shf}}>\left(\frac{m}{|q|}-\frac{|q|}{a}\right)^{-1}.
\ee
Now, 
\beer
\Phi(0)&=&F-DR^2\nn\\
&=& \left[\left(\frac{m}{q}\right)^2-1\right]^{-{\shf}}\sec^{-1}
  \left[ \left(\frac{m}{q}-\frac{q}{a}\right) \left[\left(\frac{m}{q}\right)^2-1\right]^{-{\shf}}\right] \nn\\
& & +e^{{\shf}\omega(a)} \left(\frac{m}{q}-\frac{q}{a}\right)^{-1} \left(\frac{2m}{q}-\frac{q}{a}\right)^{-1} -\left(\frac{m}{q}-\frac{q}{a}\right)^{-1}\left(\frac{2m}{q}-\frac{q}{a}\right)^{-1}\nn\\
&>& \left[\left(\frac{m}{q}\right)^2-1\right]^{-{\shf}}-\left(\frac{m}{q}-\frac{q}{a}\right)^{-1}\left(\frac{2m}{q}-\frac{q}{a}\right)^{-1}\nn\\
&>& \left(\frac{m}{|q|}-\frac{|q|}{a}\right)^{-1} \left[\left(\frac{m}{|q|}-\frac{|q|}{a}\right)^{-1}-\left(\frac{2m}{|q|}-\frac{|q|}{a}\right)^{-1}\right]\nn\\
&>&0,\nn
\eer
where we have used equation (\ref{e:mess}) on line four.  Thus the solution is non-singular and has no horizons.

A straightforward calculation shows that $p'(0)=0$, and therefore the solution is regular at the origin.

The charge density and electrostatic potential turn out to be 
\beer
\kappa \sigma&=&\pm2D\left[\left(\frac{m}{q}\right)^2-1\right]^{{\shf}}
   \sec\left[\left(\left(\frac{m}{q}\right)^2-1\right)^{{\shf}}\Phi(r)
   \right]\nn\\
   \noalign{\vskip 6pt}
& &\qquad .\left\{3\left(1-\frac{r^2}{R^2}\right)^{{\shf}}+
   Dr^2\left[\left(\frac{m}{q}\right)^2-1\right]^{{\shf}}\tan\left[
   \left(\left(\frac{m}{q}\right)^2-1\right)^{{\shf}}\Phi(r)\right]
   \right\}\nn\\
   \noalign{\vskip 6pt}
\phi&=&\frac{m}{q} \pm \left[\left(\frac{m}{q}\right)^2-1\right]^{{\shf}}
   \sec\left[\left(\left(\frac{m}{q}\right)^2-1\right)^{{\shf}}\Phi(r)
   \right]\nn
.\eer

Thus, the charge density does not change sign throughout the distribution of matter, and 
\beer
3\kappa p+\kappa\rho&=&2D^2r^2\left[\left(\frac{m}{q}\right)^2-1\right]\sec^2\left[
   \left(\left(\frac{m}{q}\right)^2-1\right)^{{\shf}}\Phi(r)\right]\nn\\
   \noalign{\vskip 6pt}
& &\qquad+12D\left[\left(\frac{m}{q}\right)^2-1\right]^{{\shf}}\sec\left[
   2\left(\left(\frac{m}{q}\right)^2-1\right)^{{\shf}}\Phi(r)\right]\left(
   1-\frac{r^2}{R^2}\right)^{{\shf}}\nn\\
   \noalign{\vskip 6pt}
&=&\pm\left(\phi-\frac{m}{q}\right)e^{-{\shf}\omega}\kappa\sigma \nn\\
   \noalign{\vskip 6pt}
&=&\frac{de^{{\shf}\omega}}{d\phi}\kappa\sigma \nn
,\eer
in agreement with equation (\ref{e:agam}).

Finally, turning to the energy conditions, we can see from the above calculation that $3p+\rho\geq 0$ and so the strong energy condition holds, while

\[
\kappa p+\kappa\rho=\frac{2}{R^2}+2 \sqrt{-B} D\sec \left[\sqrt{-B}\Phi(r)\right]\csc\left[\sqrt{-B}\Phi(r)\right]\left(1-\frac{r^2}{R^2}\right)^{\shf}\geq 0,
\]
so the weak energy condition also holds.

\subsection{Class II }

For this solution to be real and have no horizons we must have 
\[ -\frac{1}{A^2}<\Phi(r)<0.\]
Since $\Phi$ is an increasing function of $r$ , this will be true as long as
\[ \Phi(a)<0 \qquad\qquad and \qquad\qquad  \Phi(0)>-\frac{1}{A^2}.\]
The first of these is trivially true, while the second is equivalent to 
\[ \left(\frac{m}{q}-\frac{q}{a}\right)^{-1}\left(\frac{m}{q}-\frac{q}{2a}\right)^{-1}\left[1-\left(\frac{m}{q}\right)^2-\frac{m}{2a} +\frac{q^2}{2a^2} -e^{{\shf} \om(a)}\right]\exp\left[-2\left(\frac{m}{q}-\frac{q}{a}\right)^{-2}e^{\om(a)}\right]>-1.\]
A short calculation shows that this holds if
\[ \left[\frac{ 1-\left(\frac{m}{q}\right)^2-\frac{m}{2a}+\frac{q^2}{2a^2} -e^{{\shf}\om(a)} }{1-\left(\frac{m}{q}\right)^2-\frac{m}{2a}+\frac{q^2}{2a^2} -e^{\om(a)} }\right]\exp\left[-2\left(\frac{m}{q}-\frac{q}{a}\right)^{-2}e^{\om(a)}\right]<1.\]
Since both of the factors on the left-hand side are less than 1, this inequality holds. Thus the solution has no horizons.

Differentiating the pressure, it can be shown that $p'(0)=0$ and the solution is therefore regular at the origin.

Finally
\beer
\kappa p+\kappa\rho &=& \frac{2}{R^2}-\frac{2D}{\Phi(r)}\left[-\ln\left(-A^2\Phi(r)\right)\right]^{-1}\left(1-\frac{r^2}{R^2}\right)^{\shf} \nn\\
&\geq&0\nn\\
3\kappa p+\kappa\rho &=&\frac{D}{\Phi(r)^2}\left[-\ln\left(-A^2\Phi(r)\right)\right]^{-1}\left[Dr^2-6\Phi(r)\left(1-\frac{r^2}{R^2}\right)^{\shf}\right] \nn\\
&\geq&0.\nn
\eer

We conclude that both the weak and strong energy condition (as well as condition (vi)) hold for this solution.

\section{Conclusions}
Functional relationships between the gravitational and electric potentials can take many forms. In the presence of matter pressure, these relationships can go beyond the original quadratic form introduced by Weyl, and can give rise to matter distributions that satisfy many physically reasonable criteria. In addition these fields satisfy a relation between their matter, electric field and charge densities.  This generalises the fact that the active gravitational matter density and charge density of a Majumdar-type field are numerically equal.  For Majumdar-type fields the total mass and charge as measured at infinity are equal, while bounded non-singular perfect fluids with non-vanishing pressure cannot generate such fields.

The author would like to thank Petros Florides, under whose supervision most of the above work was carried out, and the referees for many helpful comments and suggestions.

\end{document}